\documentclass[aps,prx,twocolumn,superscriptaddress,longbibliography]{revtex4-2}
\usepackage{amsfonts, amsmath, amssymb}
\usepackage{graphicx}
\usepackage{dcolumn}
\usepackage{bm}
\usepackage[normalem]{ulem}
\usepackage{color}
\usepackage[utf8]{inputenc}
\usepackage[colorlinks,citecolor=blue,linkcolor=blue,urlcolor=blue]{hyperref}
\usepackage{braket}
\usepackage[caption=false]{subfig}
\usepackage{color}
\usepackage{soul}
\usepackage{mathtools}
\usepackage{float}
\newcommand{\be}{\begin{equation}}
\newcommand{\ee}{\end{equation}}
\newcommand{\ba}{\begin{eqnarray}}
\newcommand{\ea}{\end{eqnarray}}

\include{physics}

  {\left\lbrace\begin{array}{@{}l@{}}}%
  {\end{array}\right.}

\usepackage{graphicx}
\usepackage{dcolumn}
\usepackage{bm}
\usepackage{tabularx}
\usepackage{booktabs} 
\usepackage{float}
\usepackage{lipsum, babel}

\graphicspath{ {./Figures/}}

\begin{document}

\preprint{APS/123-QED}

\title{Breakdown of the topological protection by cavity vacuum fields \\ in the integer quantum Hall effect}

\author{Felice Appugliese}
\affiliation{ETH Zürich, Institute of Quantum Electronics, Auguste-Piccard-Hof 1, Zürich 8093, Switzerland }
\email{felicea@phys.ethz.ch, jerome.faist@phys.ethz.ch}

\author{Josefine Enkner}
\affiliation{ETH Zürich, Institute of Quantum Electronics, Auguste-Piccard-Hof 1, Zürich 8093, Switzerland }
\author{Gian Lorenzo Paravicini-Bagliani}
\affiliation{ETH Zürich, Institute of Quantum Electronics, Auguste-Piccard-Hof 1, Zürich 8093, Switzerland }
\altaffiliation{Present address: University of Strasbourg, CNRS, ISIS, 67000 Strasbourg, France.}
\author{Mattias Beck}
\affiliation{ETH Zürich, Institute of Quantum Electronics, Auguste-Piccard-Hof 1, Zürich 8093, Switzerland }
\author{Christian Reichl}
\affiliation{ETH Zürich,Laboratory for Solid State Physics, Zürich 8093, Switzerland}
\author{Werner Wegscheider}
\affiliation{ETH Zürich,Laboratory for Solid State Physics, Zürich 8093, Switzerland}
\author{Giacomo Scalari}
\affiliation{ETH Zürich, Institute of Quantum Electronics, Auguste-Piccard-Hof 1, Zürich 8093, Switzerland }
\author{Cristiano Ciuti}
\affiliation{Universit\'e de Paris, Laboratoire Mat\'eriaux et Ph\'enom\`enes Quantiques (MPQ), CNRS-UMR 7162, France }
\author{J\'{e}r\^{o}me Faist}
\affiliation{ETH Zürich, Institute of Quantum Electronics, Auguste-Piccard-Hof 1, Zürich 8093, Switzerland }

\begin{abstract}
The control of the electronic properties of materials via the vacuum fields of cavity electromagnetic resonators is one of the emerging frontiers of condensed matter physics.
We show here that the enhancement of vacuum field fluctuations in subwavelength split-ring resonators dramatically affects arguably one of the most paradigmatic quantum protectorates, namely the quantum Hall electron transport in high-mobility 2D electron gases. The observed breakdown of the topological protection of the integer quantum Hall effect is interpreted in terms of a long-range cavity-mediated electron hopping where the anti-resonant terms of the light-matter coupling finally result into a finite resistivity induced by the vacuum fluctuations. The present experimental platform can be used for any 2D material and provides new ways to manipulate electron phases in matter thanks to vacuum-field engineering. 
\end{abstract}

\maketitle

One of the most intriguing aspects of quantum field theories is the description of empty space as permeated by electromagnetic vacuum fluctuations~\cite{loudon2000quantum}. Energy conservation forbids any process that would lead to net energy extraction from such states, implying in particular that such vacuum fields cannot be detected by direct absorption. Nevertheless, there are many experimental evidences of their existence~\cite{milonni1994the}: spontaneous emission, Lamb shift, Casimir effect~\cite{casimir1,Munday:2009} can only be explained by invoking the role of vacuum fields.  More recently, technological developments of both laser sources and optical nano-cavities in the strong light-matter coupling offer a new perspective about vacuum fields, one in which those fluctuations can be sensed directly~\cite{riek2015direct,benea2019electric} and used to engineer new  properties of matter~\cite{forn2019ultrastrong,kockum2019ultrastrong,GarciaVidal2021,Hubener:2020fm,orgiu2015conductivity} without illumination. Optical excitations in the strong light-matter coupling regime, the so-called polaritons, are also used as sensitive probes of many-body states such as fractional quantum Hall states~\cite{Smolka:2014iu}, Wigner Crystals~\cite{smolenski2021} or of correlated photon phases~\cite{goblot2019nonlinear}. Confining the light on a strongly subwavelength scale is one of the key elements that enabled the achievement of record high interaction strengths~\cite{Todorov:2010p1832,scalari2012ultrastrong,bayer2017terahertz} in which the anti-resonant terms of the Hamiltonian play an important role~\cite{ciuti2005quantum}. Cavity-controlled superconductivity~\cite{Schlawin:2019jw,thomas2019}, long-range ferroelectric order~\cite{Ashida2020,thomas10.1021/acs.nanolett.1c00973}, cavity-mediated superradiance~\cite{GarciaVidal2021} based on cavity-mediated electron or dipole interactions have been recently investigated. 
Experimentally, however, unambiguous identification of modified equilibrium properties of solid-state quantum phases of matter through vacuum fields remains an open challenge. 
We recently demonstrated the role of Landau polaritons~\cite{hagenmuller2010ultrastrong} in controlling the DC bulk magneto-transport~\cite{paravicini2019magneto} in a semiconductor electronic gas, even in absence of external illumination. The amplitude of the Shubnikov-de Haas oscillations was substantially modified by the presence of a THz resonator embedding the Hall bar. However, as a change in the shape of the oscillations in DC magneto-resistivity is sensitively linked to variations of scattering processes, the phenomenology is not universal and every sample is unique. Instead, we propose to investigate here transport in the integer quantum Hall regime~\cite{KlitzingPRL1980}, where the topological protection of the edge currents results in a quantized resistance and where the effect of vacuum fields can be unambiguously evidenced. 
At low magnetic fields it is possible to describe transport in the presence of impurity scattering only as a broadening of the Landau levels' density of states. In the high magnetic fields instead the random impurity potential causes the localization of electrons.
When the Fermi energy lies around a minimum between consecutive Landau levels, both the longitudinal magneto-conductivity $\sigma_{xx}$ and magneto-resistivity $\rho_{xx}$ simultaneously vanish. In order to describe this non trivial effect, is necessary to consider the role of the states at the sample edge.
\begin{figure*}
 \includegraphics[width=\textwidth]{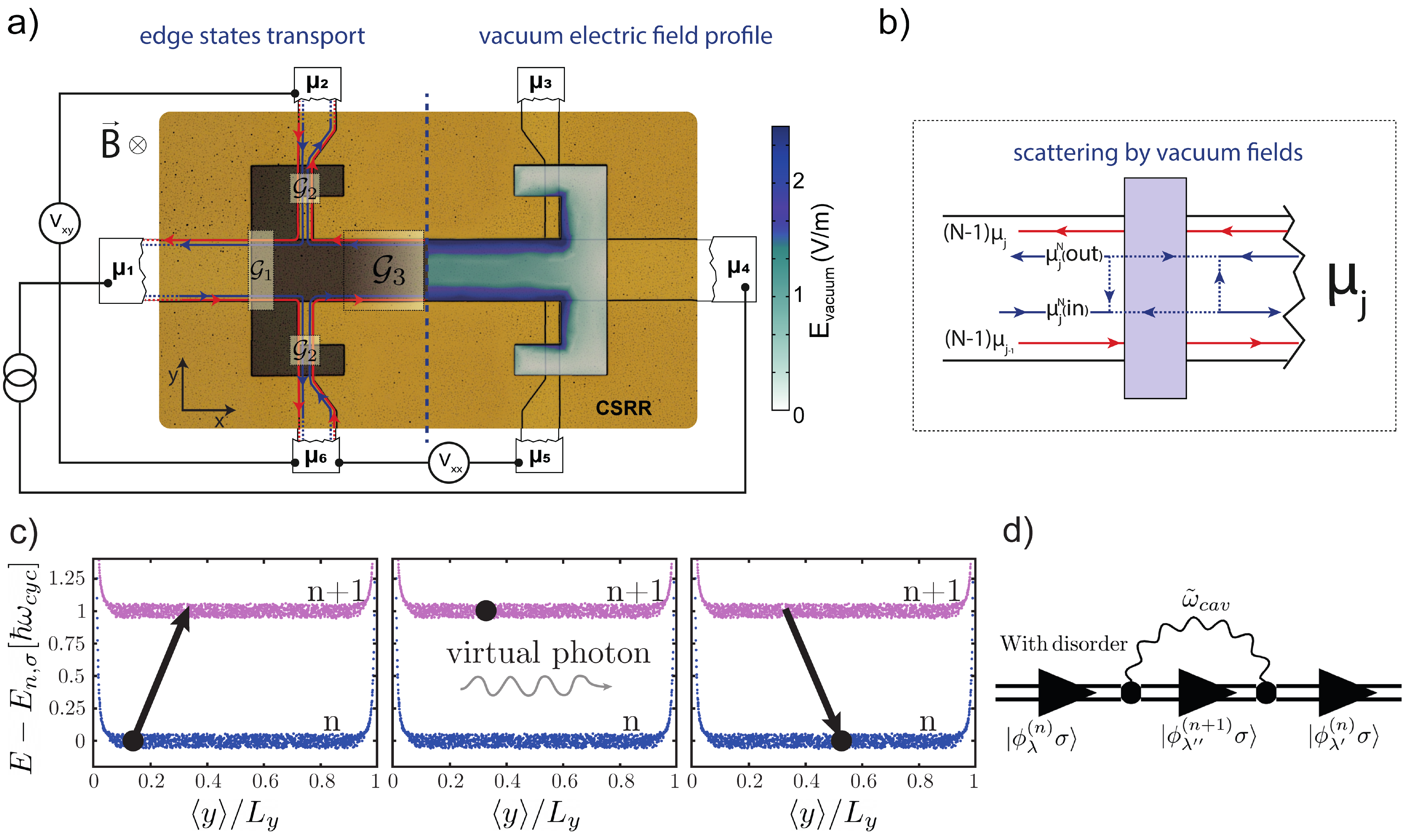}
  \caption{(a) Rendering of a complementary THz split-ring 140GHz resonator embedding a Hall bar with its electrical contacts. The  resonator is defined by the exterior metallic gold layer, while the 2D electron gas is inserted within the region without metal. Without illumination and in the linear transport regime, only vacuum fields of the confined electromagnetic modes can affect the electronic transport. (b) The edge conduction channels for the 2D electron gas are also sketched. 
  (c) Schematic representation of a cavity-mediated electron hopping process in the normally insulating region due to the electromagnetic resonator vacuum fields. An electron in a disordered eigenstate belonging to the $n$ Landau band can be promoted to the $(n+1)$ band with the emission of a virtual cavity photon. Such anti-resonant process can occur thanks to the counter-rotating terms of light-matter interaction in the cavity system that become relevant in the ultra-strong light matter coupling regime. The reversed process couples the intermediate state to a different final state for the electron. (d) A diagrammatic representation of the same cavity-mediated hopping process (the electron spin $\sigma$ is conserved), describing the hopping between state $\lambda$ in the $n$ disordered Landau band and the state $\lambda'$ in the same $n$ band via the intermediate state $\lambda''$ in the $(n+1)$ band.
  } 
  \label{fig:schematics}
\end{figure*}
The simplest picture that best describes the physics of this regime is built on the work of Landauer~\cite{landauer1970electrical} and B{\"u}ttiker~\cite{buttiker1988absence}.
The key point is considering the bulk as insulating, due to the localization of electrons in a fluctuating spatial potential, while the edges are conducting.
The current flows through a discrete number of one-dimensional edge channels from one contact to the following in clockwise direction, each of them contributing with $\frac{e^2}{h}$ to the conductance and exhibiting zero longitudinal resistance. 
This system is a prototypical topological insulator~\cite{Hasan_RevModPhys2010}, where the edge currents are chiral, traveling in opposite directions at the two opposite edges of the sample. As a result, an electron in an edge channel cannot back scatter unless it is scattered on the other edge of the sample. This is true for potentials which vary slowly over a cyclotron radius but rapidly over the sample dimensions. The impurity potential can be strong, but it still does not produce backscattering when it is short range~\cite{buttiker1988absence}. For this reason, the topological protection of the quantum Hall effect is robust against local perturbations such as a static disorder. A significant change in the integer quantum Hall phenomenology can only be explained by invoking the non-local nature of the cavity vacuum fields, as the nonlocality is the Achille's heel of topological order.
Here we report a comprehensive set of experimental data revealing significant and intriguing modifications of the integer quantum Hall effect in Hall bars immersed in an electronic resonator where the vacuum field fluctuations are enhanced by their confinement into a strongly subwavelength volume. Fig.~\ref{fig:schematics} shows schematically the geometry of our experiment~\cite{paravicini2019magneto}. A Hall bar of width 40$\mu$m, fabricated on a high-mobility two-dimensional electron gas (See Supplementary Material for details) is located in the spatial gap of a complementary metallic resonator~\cite{OpexComplementaryChen2007,maissen2014ultrastrong} with a resonance at 140 GHz. The longitudinal and Hall transverse resistance are measured within the gap of the resonator. The geometry of the resonator is such that the vacuum field fluctuations $\mathcal{E} = \sqrt{\hbar \omega_{\rm cav}/2\epsilon_0 \epsilon_s V_{eff}}=1 V/m$ are strongly enhanced as compared to the free space, where the amplitude would be of the order of $\mathcal{E}=65 mV/m$ in that frequency range, and lead to light-matter collective Rabi frequency   $\tilde{\Omega}_R \simeq 0.3 \omega_{\rm cav}$, where $\omega_{\rm cav}$ is the cavity mode angular frequency, as indeed observed in optical characterization experiments~\cite{paravicini2019magneto,paravicini2017gate}. As shown in Fig.~\ref{fig:schematics} a), this vacuum field is fairly homogeneous in the center of the resonator but increases towards the edges. 
As displayed schematically in Fig.~\ref{fig:schematics}, the electron transport  in the integer quantum Hall effect proceeds via edge states that are protected against backscattering by the lifting of the time-reversal symmetry provided by the magnetic field. As shown in a recent work~\cite{CC_detailed_theory}, in the presence of disorder a  cavity-mediated long-range hopping can be achieved via the exchange of a virtual cavity photon (see Fig.~\ref{fig:schematics}). For an electron occupying a disordered eigenstate $\vert \phi^{(n)}_{\lambda}\rangle$ in the $n$-th Landau band, the coupling to another disordered eigenstate $\vert \phi^{(n)}_{\lambda'}\rangle$ in the same Landau band occurs via light-matter coupling to an intermediate state, consisting of an electron in the state $\vert \phi^{(n+1)}_{\lambda''} \rangle$ in the $(n+1)$-th band and one cavity photon. At the lowest order in perturbation theory, the effective coupling between disordered states $\lambda$ and $\lambda'$ in the same $n$ Landau band is given by the expression 
\begin{equation}
    \Gamma^{(n)}_{\lambda,\lambda'} = \sum_{\lambda''}\frac{\tilde{g}^{(n,n+1)}_{\lambda,\lambda''} \tilde{g}^{(n,n+1)*}_{\lambda',\lambda''}}{\epsilon_{n,\lambda}-\epsilon_{n+1,\lambda''}-\hbar \tilde{\omega}_{\rm cav}}\, 
    \label{eq:Gamma_lambda_lambda_prime}
\end{equation}
and can be thought as a cavity-mediated hopping. Importantly, note that this intermediate virtual process is due to the counter-rotating terms of the light-matter interaction and is quantified by the single-electron vacuum Rabi frequency $\tilde{g}^{(n,n+1)}_{\lambda,\lambda'}$, which depends on the vacuum Rabi frequency $g$ without disorder, the wavefunctions of the disordered eigenstates and is softened by diamagnetic renormalization, as detailed in Ref.~\cite{CC_detailed_theory}. The number of intermediate states is equal to the Landau degeneracy $N_{\rm deg} = N_{\rm el}/\nu$, where $N_{\rm el}$ is the number of electrons and $\nu$ the filling factor.   The denominator of the expression consists of the energy penalty associated to the excitation of the virtual intermediate state described above. Such cavity-mediated hopping involve all the disordered eigenstates, affecting both edge and bulk states. For the quantitative impact of the process, a crucial role is played by the collective vacuum Rabi frequency $\tilde{\Omega}_R = \tilde{g} \sqrt{N_{\rm el}}$, where the tilde indicates the diamagnetic renormalization~\cite{CC_detailed_theory}. Note that, given the anti-resonant nature of the process, additional electromagnetic modes with higher frequencies can contribute in the same qualitative way and re-inforce quantitatively the effective cavity-mediated hopping. In addition, the strong electromagnetic field gradients present at the edge of the metal resonator, as apparent in Fig.~\ref{fig:schematics}(a) will play the same role impurity disorder in enabling cavity-mediated hopping~\cite{CC_detailed_theory}.
\begin{figure*}
  \includegraphics[width=\textwidth]{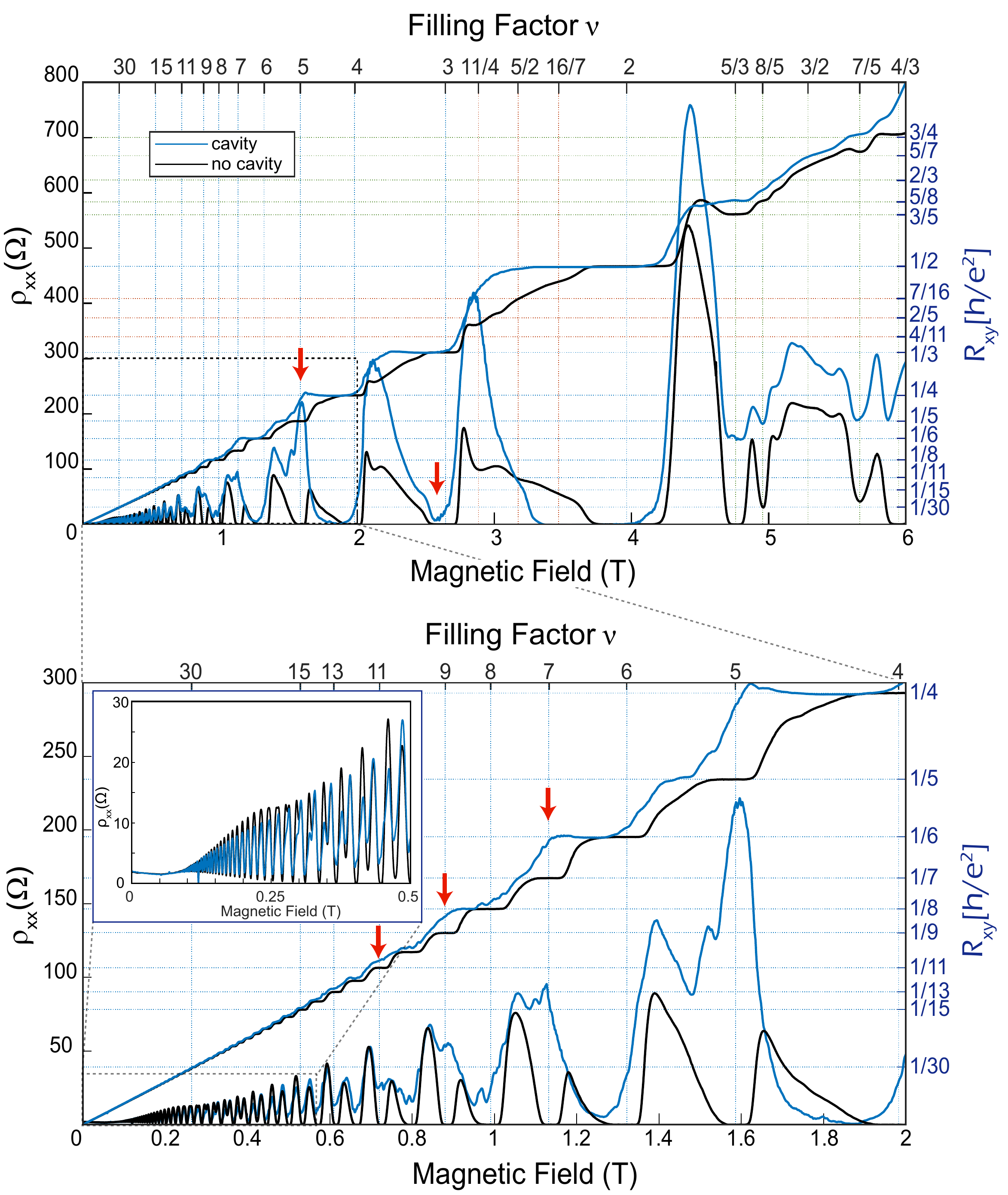}
  \caption{Longitudinal and transverse resistance for a reference Hall bar (black lines) and for a cavity embedded hall bar (blue lines). The region from $0$ to $0.5$ T is also shown in the inset for better visibility, in particular showing that both the longitudinal and transverse resistance with and without cavity converge to the same values in the limit of zero magnetic field.}
  \label{fig:vxx_vxy_sample1}
\end{figure*}
The transverse and longitudinal magnetoresistances of a reference sample and of a Hall bar coupled to  a cavity are compared in Fig.~\ref{fig:vxx_vxy_sample1}. The sample exhibits a mobility of $16\times 10^6 \text{cm}^2/\text{Vs}$ and a density $2\times 10^{11} \text{cm}^{-2}$ and is kept in the dark in a dilution fridge at the nominal temperature of 10mK, while the electron temperature is estimated to be close to 50 mK. Both samples are measured in the same cooling run and are physically located onto the same chip. 
While for the reference sample (no cavity), all the integer quantum Hall plateaus are well developed, showing zero longitudinal resistance down to $\nu=11$, the sample with cavity displays a  strong deviation from the integer  quantization for all the odd plateaus. At the same time, as is apparent in the inset of  Fig.~\ref{fig:vxx_vxy_sample1}, the resistance at zero magnetic field is essentially unchanged and the transport features associated with the fractional quantum Hall regime are  only weakly affected. This is important because it shows that the mobility and overall quality of the sample are unchanged by the complementary resonator (indeed, the metallic gold layer defining the resonator is deposited around the electronic sample and well separated from it) and also remarkable because the  fractional quantum Hall features are usually the most fragile with respect to perturbations~\cite{StormerRevModPhys.71.S298}.
Note also that, as shown in the inset of Fig.~\ref{fig:vxx_vxy_sample1}, at low magnetic fields ($B<0.3$T) the cavity sample displays a reduction of the amplitude of Shubnikov-de Haas oscillations with respect to the reference sample, as already evidenced and discussed in our previous study~\cite{paravicini2019magneto}.
The experiments shown in Fig.~\ref{fig:vxx_vxy_sample1} were repeated for a number of temperatures up to 1 K, allowing us to extract the activation energies of these transport features. The results are reported in Fig.~\ref{fig:activation_energy_sample1} and confirm the picture seen so far: while the activation energy of the reference sample is of the order of the Zeeman splitting, the one in presence of  the cavity is dramatically reduced.  The very same activation energy analysis performed on the fractional Hall states shows that the presence of the cavity has only a very weak effect on the activation energy and therefore on the transport gap. This is consistent with the fact that fractional Hall states couple extremely  weakly to  electromagnetic fields~\cite{Kukushkin:2002tn}, which is essentially a consequence of Kohn's theorem~\cite{KohnPhysREv1962}.
\begin{figure*}
  \includegraphics[width=\textwidth]{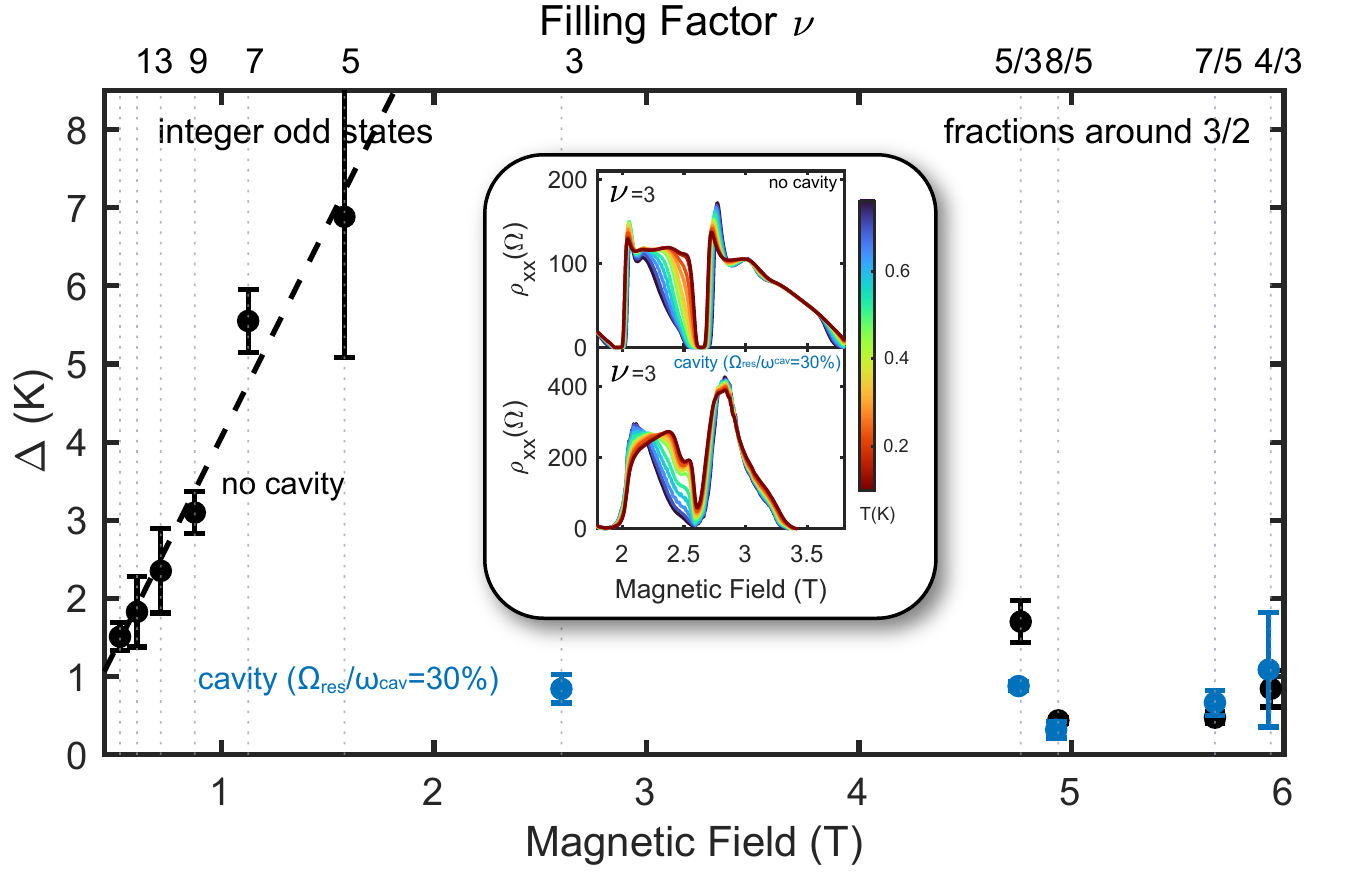}
  \caption{Comparison of the transport activation energies extracted from temperature-dependent measurements of the longitudinal resistance at integer odd filling factors for the reference Hall bar (no cavity, black) and for the cavity sample (blue). The cavity produces a dramatic decrease of the activation energy. Fractional filling factors appear to be largely immune from the presence of the cavity. Indeed, fractional quantum Hall states couple very weakly to the electromagnetic field despite being typically more fragile to the effects of disorder and a finite temperature. (Inset) Longitudinal resistance as a function of temperature, in the range $50$ mK-$1$K for magnetic field close to the $\nu =3$ plateau for the reference (top) and cavity Hall bar (bottom).}
  \label{fig:activation_energy_sample1}
\end{figure*}
Cavity-mediated hopping is expected to conserve the electron spin. Moreover, for edge states the cavity-mediated hopping vanishes when the difference between the edge state energy and the bulk is much larger than the bandwidth of the disordered Landau band~\cite{CC_detailed_theory}. 
However, due to the fact that in GaAs the Zeeman spin splitting is significantly smaller than the cyclotron splitting between Landau levels, the odd integer filling factor plateaus are more fragile, since they are protected by the smaller spin-splitting gap~\cite{girvin_yang_2019}. 
Indeed, in the absence of a cavity, the relevant energy gap for even integer quantum Hall states is the cyclotron energy $E_{cyc}=\hbar\frac{e B_{\perp}}{m_{\star}}$, where $B_{\perp}$ is the projection of the magnetic field perpendicular to the plane of the 2DEG. On the other hand, the energy gap protecting the odd states is the Zeeman energy $E_{\text{Ze}}=g_{\star} \mu_{\rm B}B_{\text{tot}}$, where $B_{\text{tot}}$ is the total magnetic field applied to the sample, $\mu_{\text{B}}$ is the Bohr magneton and $g_{\star}$ is the effective electron g-factor. Due to electron-electron interaction, $g_{\star}$ can become bigger than $2$, namely the value for a free electron, and the Zeeman energy can become a significant fraction of the cyclotron energy ($\sim$0.2 in the samples investigated at normal magnetic field incidence)~\cite{Nicholas:1988vl}.
It is possible to tune the ratio of Zeeman and cyclotron energy, simply because the first depends on the total magnetic field applied to the sample, while the second only depends on the perpendicular component, as shown in Fig.~\ref{fig:Zeeman_ratio} a). The perpendicular component of the magnetic field is equal to $B_{\perp}=B_{\text{tot}}\cos(\theta)$, where $\theta$ is the angle between the total applied magnetic field and the normal to the plane of the quantum well.  In the considered experiment, we were able to vary $\theta$ from $0$ to $50^{\circ}$ by rotating the sample during the measurement run, so the ratio can be enhanced by a factor $1.5$ with respect to the $\theta = 0$ value.
\begin{figure*}
  \includegraphics[width=\textwidth]{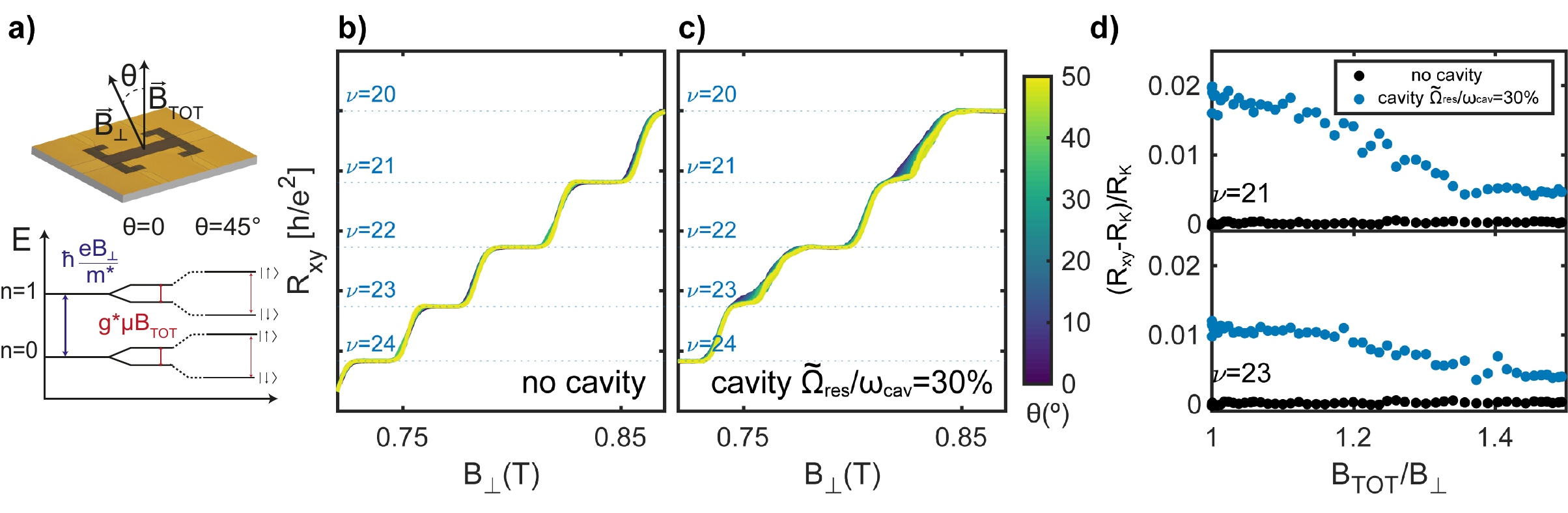}
  \caption{Transverse resistance as a function of magnetic field for different values of the finite tilting angle $\theta$ that allows us to tune the ratio between the Zeeman and cyclotron energy splitting, as depicted in panel (a). Panel (b) and (c) report respectively the transverse resistance without and with cavity as a function of the magnetic field component $B_{\perp}$ (keeping $B_{\text{tot}}$ constant). Panel (d) displays the deviation of the transverse resistance from the quantized integer value as a function of ratio $B_{\text{tot}}/B_{\perp}$ for the two odd integer filling factors.}
  \label{fig:Zeeman_ratio}
\end{figure*}
In Fig.~\ref{fig:Zeeman_ratio} b) and c), the transverse resistance from filling factors $\nu=20$ to $\nu=24$ is reported for a second sample (D170209B) that exhibits a larger electron density ($n_s = 4\times 10^{11} cm^{-2}$ and a similarly high mobility). The colormap from dark to bright shows the evolution of the curves for different values of the tilting angle. It is evident that there is a larger change in the cavity traces compared to the reference Hall bar Fig.~\ref{fig:Zeeman_ratio} d), we report the deviation from the integer quantized value as a function of the ratio between total and perpendicular magnetic field.
This deviation decreases when the ratio increases, confirming that increasing the energy gap indeed weakens the cavity-induced scattering.
The experiments show that vacuum fluctuations, strongly enhanced in the gap of a metallic resonator, generate a long-range hopping that breaks the quantization of the resistance. The quantitative characterization of this phenomenon can be conveniently performed adopting the formalism based on the Landauer-B{\"u}ttiker edge state picture~\cite{szafer1991network}. As shown schematically in Fig.~\ref{fig:schematics}, in this theoretical framework, the deviation from quantization is interpreted by a finite transmission $t_i$ of the highest populated edge state of section $i$ of the conductor due to scattering of a fraction $(1-t_i)$ to the other edge. The strength of the scattering for the edge state $\nu$ is related to length $L_i$ and the width $w_i$ of the conductor to a resistivity $\rho_{xx}^{\nu}$  by the relation $t_i = 1/(1+\rho_{xx}^{\nu}L_i/w_i)$. This model was successfully used to explain how the longitudinal resistance in the region between quantized plateaus depended on the geometry of the voltage probes~\cite{mceuen1990new}.  
In our case, we will interpret $\rho_{xx}^{\nu}$ as the "resistivity" originating from the vacuum fluctuations which will concern only the fraction of the conductor exposed to this field. Such an interpretation, however, {\em requires} that the rest of the conductor is in the quantum Hall regime and as a result this analysis will only be performed in the middle of a quantum Hall plateau. 
This model was implemented and solved for a standard Hall bar with 6 contacts, as detailed in the supplementary material section. The experimental inputs are the three length-to-width ratios describing the interaction of the vacuum field with the current, voltage probes as well as the main part of the Hall bar  (designed by $\mathcal{G}_1,\mathcal{G}_2$ and $\mathcal{G}_3$, respectively in  Fig.~\ref{fig:schematics}). 
As shown in Fig.~\ref{fig:Buttiker_formula}, using these parameters, the measured values of the longitudinal resistances for the cavity sample are used to fit the relevant value of $\rho_{xx}^{\nu}$, while the computed value of the corresponding $R_{xy}$ is displayed along with the experimental data in Fig.~\ref{fig:Buttiker_formula}. The magnetic field values are chosen in the middle of the plateau measured on the reference sample. The excellent agreement between theory and experiment for the Hall resistance is an indication that this model represents well the experimental results. The value of $\rho_{xx}^{\nu}$ is displayed as a function of $\nu$ in  Fig.~\ref{fig:Buttiker_formula} (b) (blue dots) for values of the magnetic field corresponding to the center of the plateau of the reference sample. 
As already discussed, the effect of the vacuum field fluctuations is stronger on odd plateau that are protected by the smaller Zeeman gap. Correspondingly, the value of $\rho_{xx}^{\nu}$ for $\nu = 3..5$ is about an order of magnitude larger than for $\nu = 4..6$. The general increase of $\rho_{xx}^{\nu}$ with $\nu$ is easily understood by the dependence of the gaps in the magnetic field. As $\rho_{xx}^{\nu}$ becomes larger than 1 for $\nu = 7$ means that most of the edge state is scattered in the transverse resistance, implying that the transverse resistance has approximately the value corresponding to $\nu = 6$. To illustrate the consistency of the results, we report another set of data taken from a different resonator on the same heterostructure material, having the same shape but with different voltage leads geometry (light blue dots). The results, although not identical, still exhibit the same behaviour both qualitatively and quantitatively.
Finally, on the same graph are also reported the values for an identical resonator fabricated on a third high-mobility 2D gas (EV2124) with the same electron density but grown in a different MBE reactor and exhibiting a mobility reduced by a factor $8$. There, again, similar values are obtained showing that the scattering mechanism has a weak dependence on the disorder.   
\begin{figure*}
  \includegraphics[width=\textwidth]{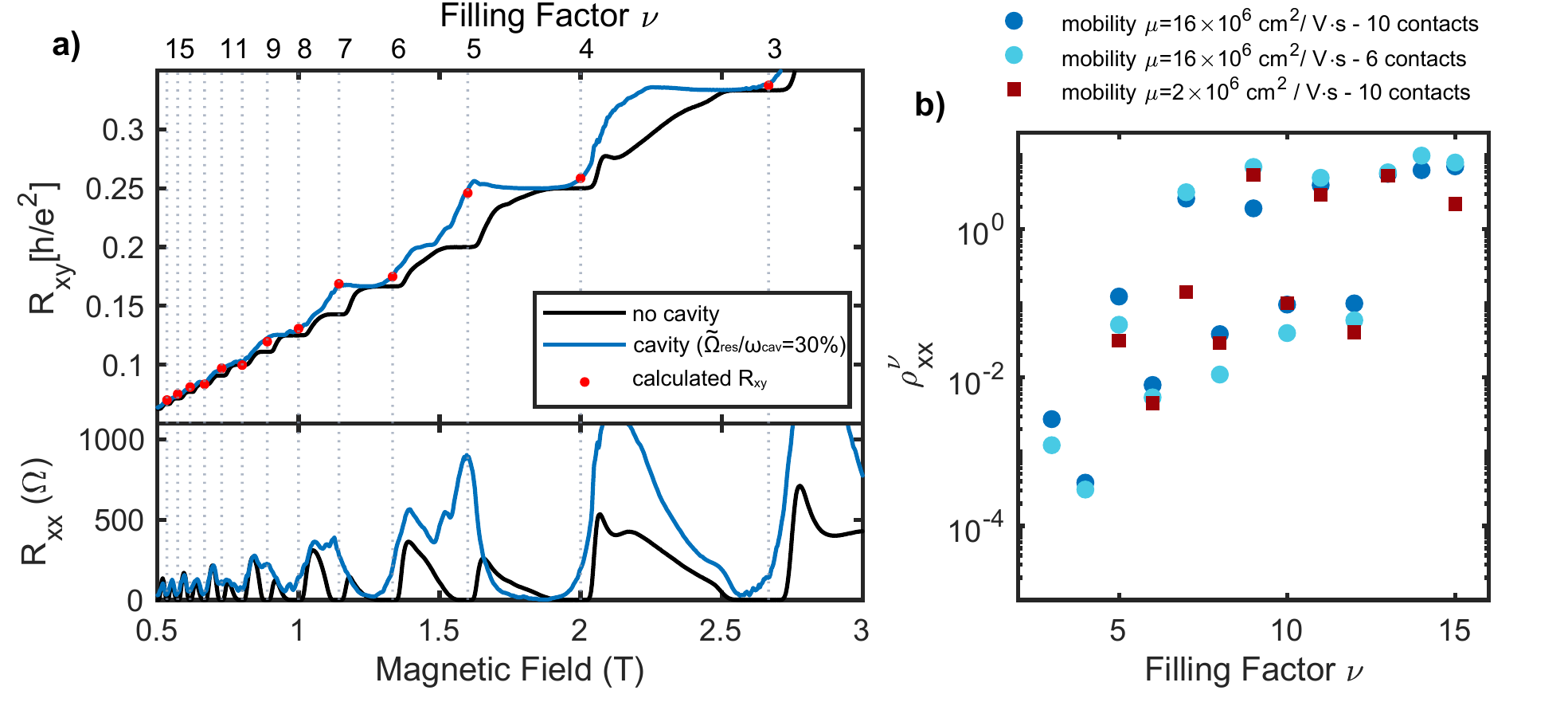}
  \caption{(a) Interpreting the effect of vacuum field by a resistivity $\rho_{xx}^{\nu}$  at integer filling factors, we have calculated the transverse resistance (red points, upper panel) from the measured longitudinal resistance. The prediction is consistent with the measured transverse resistance (blue curves). Note that for the reference sample with no cavity, the longitudinal resistance is negligible. (b) The extracted resistivity $\rho_{xx}^{\nu}$ as a function of filling factor for three different samples having very similar densities. 
 }
  \label{fig:Buttiker_formula}
\end{figure*}
The scattering mechanism arising from the coupling to the vacuum fluctuations in the cavity is expected to increase strongly with the normalized light-matter coupling strength $\tilde{\Omega}_R/{\omega_{\text{cav}}}$. As shown in Fig.~\ref{fig:coupling_dependence} a), we investigated this dependence by designing a series of three cavities with the fundamental mode having the same frequency, but exhibiting different coupling strengths $\tilde{\Omega}_R/{\omega_{\text{cav}}} = 0.17$, $0.2$ and $0.22$. As discussed in the supplementary material section, the coupling strength was computed by finite-element simulations, and confirmed by experimental measurements  on a separate sample. The heterostructure used (F150817A) exhibited similar densities and mobilities as compared to the first one used for the measurements shown in Fig.~\ref{fig:vxx_vxy_sample1}, and as expected (see the supplementary material), the longitudinal and transverse Hall measurements displayed similar loss of quantization on the odd Hall plateau as the one reported in Fig.~\ref{fig:vxx_vxy_sample1}. A slight residual parallel conduction present in the sample however reduced the range of values of $\rho_{xx}^{\nu}$ that could be extracted to values larger than $\rho_{xx}^{\nu} \approx 10^{-2} R_K$. In   Fig~\ref{fig:coupling_dependence}, we show  the values of $\rho_{xx}^{\nu}$ for $\nu = 7,9,11,13,15$. As expected, a strong increase of the resistance is observed as a function of the light-matter coupling strength $\Omega_R$, further demonstrating that the scattering originates from the vacuum field. 
\begin{figure*}
  \includegraphics[width=\textwidth]{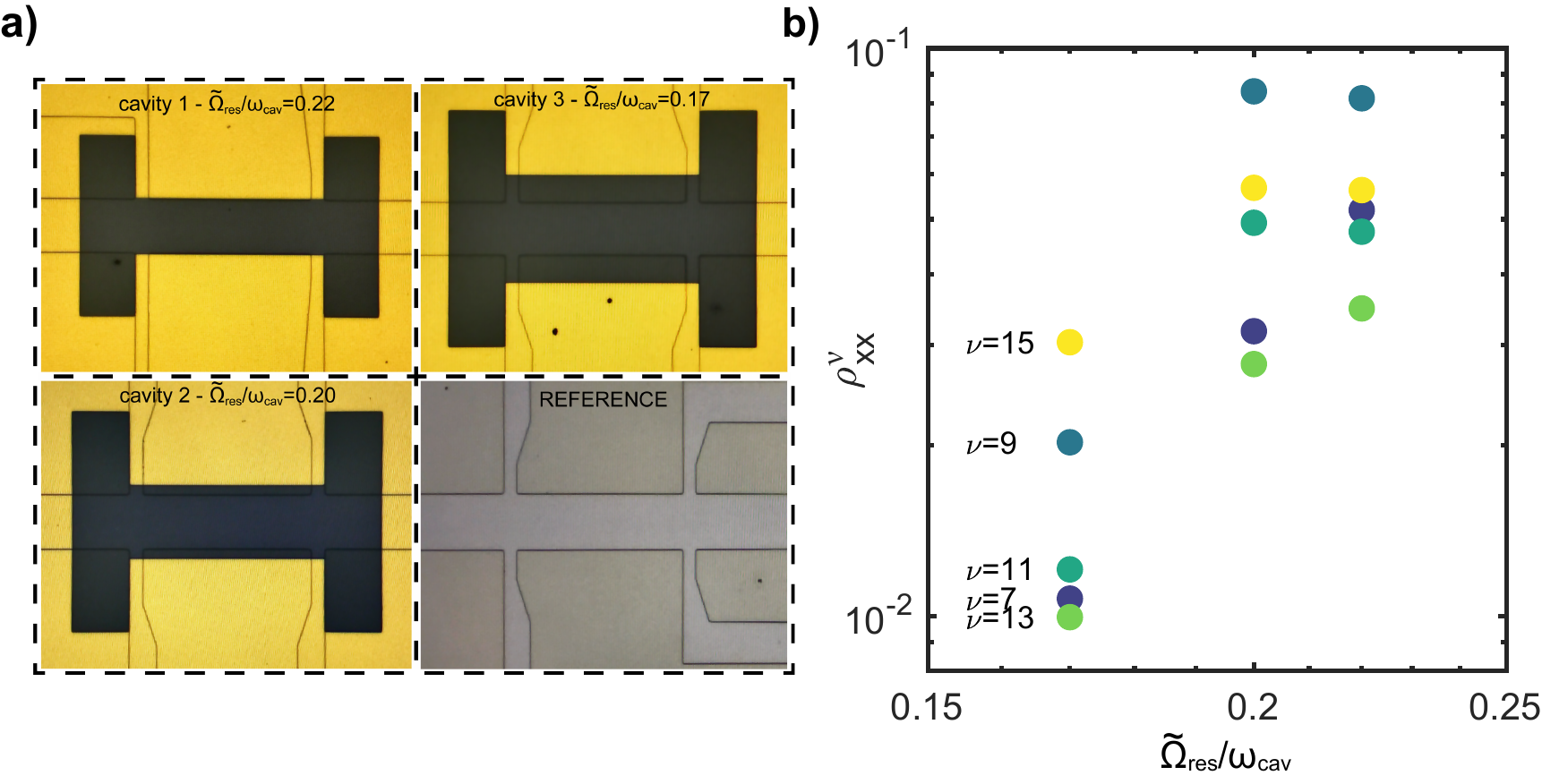}
  \caption{a) Optical micrographs of the metallic complementary resonators with their normalized coupling strength. b) Log-log plot of the resistivity $\rho_{xx}^{\nu}$ as a function of the normalized coupling strength $\tilde{\Omega}_R/{\omega_{\text{cav}}}$  for different filling factors. 
  }
  \label{fig:coupling_dependence}
\end{figure*}
As mentioned already, by only considering the fundamental resonance of the resonators we are neglecting the contribution of the higher-lying resonances. Our experimental evidence show that while these couplings should be taken into account for a more accurate quantitative estimate of the effect, the measured coupling strengths of the resonators are decreasing for higher frequency modes. In addition, the very similar geometry of the resonators that are compared imply that the respective contributions of the higher frequency modes will follow the same trend as the one of the fundamental mode. 
In conclusion, we have shown that the vacuum field in a deeply subwavelength electromagnetic resonator produces a break-down of the topological protection of the integer quantum Hall effect. The strong effect reported here on the quantum Hall plateaus and longitudinal resistance is such that one can legitimately wonder whether the much weaker vacuum fluctuations in free space might be what ultimately limits the extreme metrological precision of quantum Hall resistance standards. Indeed, in free space we expect vacuum fields about 15 times smaller than in the present split-ring resonators. Given that the vacuum-induced scattering scales as the square of the cavity-mediated hopping matrix element in (\ref{eq:Gamma_lambda_lambda_prime}), then the effect should be re-scaled by a factor $(1/16)^4 \simeq 10^{-5}$. Comparing this number with the accuracy achieved metrologically ($\sim 10^{-10}$) is challenging since those measurements are performed on much wider Hall bars and larger magnetic fields~\cite{jeckelmann2001}.
The role of vacuum fields might be also explored in the context of other topological systems such as Majorana fermions in semiconducting quantum wires. There, in contrast, finger gates~\cite{Lutchyn:2018hq} with submicron gaps that play the role of antennas and enhance vacuum fields are an essential component of these devices and may set strong limit to the topological protection offered to the quantum bits. 
From a broader perspective, the results of the present work  provide strong evidence that vacuum fluctuations can be engineered to create and/or control new electronic states of matter. The complementary split-ring resonator platform explored here for the quantum Hall regime in GaAs is very general and can be applied to any two-dimensional conductor, such as for example 2D van der Waals materials and their heterostructures. Our approach can be used to investigate for example monolayer and twisted bilayer graphene. Vacuum fields could be judiciously exploited to modify electronic localization properties, control superconductivity or other prominent condensed matter quantum phases.      
\section*{Acknowledgements}
The Authors would like to thank J. Andberger for help in the initial stage of the project, P. M\"arki for technical support and A. Imamoglu for discussions. 
The Authors  acknowledge financial
support from the ERC Advanced grant Quantum Metamaterials
in the Ultra Strong Coupling Regime (MUSiC) (grant n
340975).
The authors also acknowledge financial support from the Swiss National Science Foundation (SNF) through the National Centre of Competence in Research Quantum Science and Technology (NCCR QSIT) as well as from the individual grant $200020\_192330$.
C.C. acknowledges support from the ANR project TRIANGLE (ANR-20-CE47-0011) and from the FET FLAGSHIP Project PhoQuS (grant agreement ID no. 820392).
\bibliography{References.bib}
\cleardoublepage

\onecolumngrid
\renewcommand{\thefigure}{S\arabic{figure}}
\renewcommand{\thetable}{S\arabic{table}}
\setcounter{figure}{0}    
\section*{Supplementary material}

\subsection*{Details of the heterostructures used in the experiments}

The experiments were performed on samples processed from four different GaAs-based heterostructures. The densities and mobilities are summarized in Table \ref{tab:sample_heterostructures}.

\begin{table}[H]
    \centering
        \begin{tabular}{c c c c}
        \toprule
        Name &$n_{s}(cm^{-2})$&$\mu(cm^2)/V\cdot s$ &$Figures$ \\
        \midrule
        D170608A & $2.0\times10^{11}$ & $16\times10^6$ &2,3,5\\
        D170209B & $4.2\times10^{11}$ & $18\times10^6$ & 4\\
        F150817A & $1.9\times10^{11}$ & $12\times10^6$ & 6\\
        EV2124 & $1.9\times10^{11}$ & $2\times10^6$ & 5(b)\\
        \bottomrule
    \end{tabular}
    \caption{Values of the relevant parameters of the four heterostructures}
    \label{tab:sample_heterostructures}
\end{table}

The four heterostructures, grown in three different MBE reactors (as the prefix D,F,EV indicates) have the following properties:
\begin{itemize}
\item D170608A is a double side doped, 27 nm wide GaAs/AlGaAs quantum well, with a 110/120 nm spacer.
\item D170209B is a double side doped, 27 nm wide GaAs/AlGaAs quantum well with a 60nm spacer.
\item F150817A is a double side doped, 27 nm wide GaAs/AlGaAs quantum well with a 120 nm spacer. 
\item EV2124 is a triangular well 90 nm deep, with a Si $\delta$-doping layer 50 nm below the surface. 
\end{itemize}
Note that the values of mobilities and densities have been measured on the reference Hall bar at the temperature of approximately $50$ mK without external illumination. 
\subsection*{Modeling the effect of the vacuum field on the resistance quantization}
\begin{figure}[htb]
  \includegraphics[width=\textwidth]{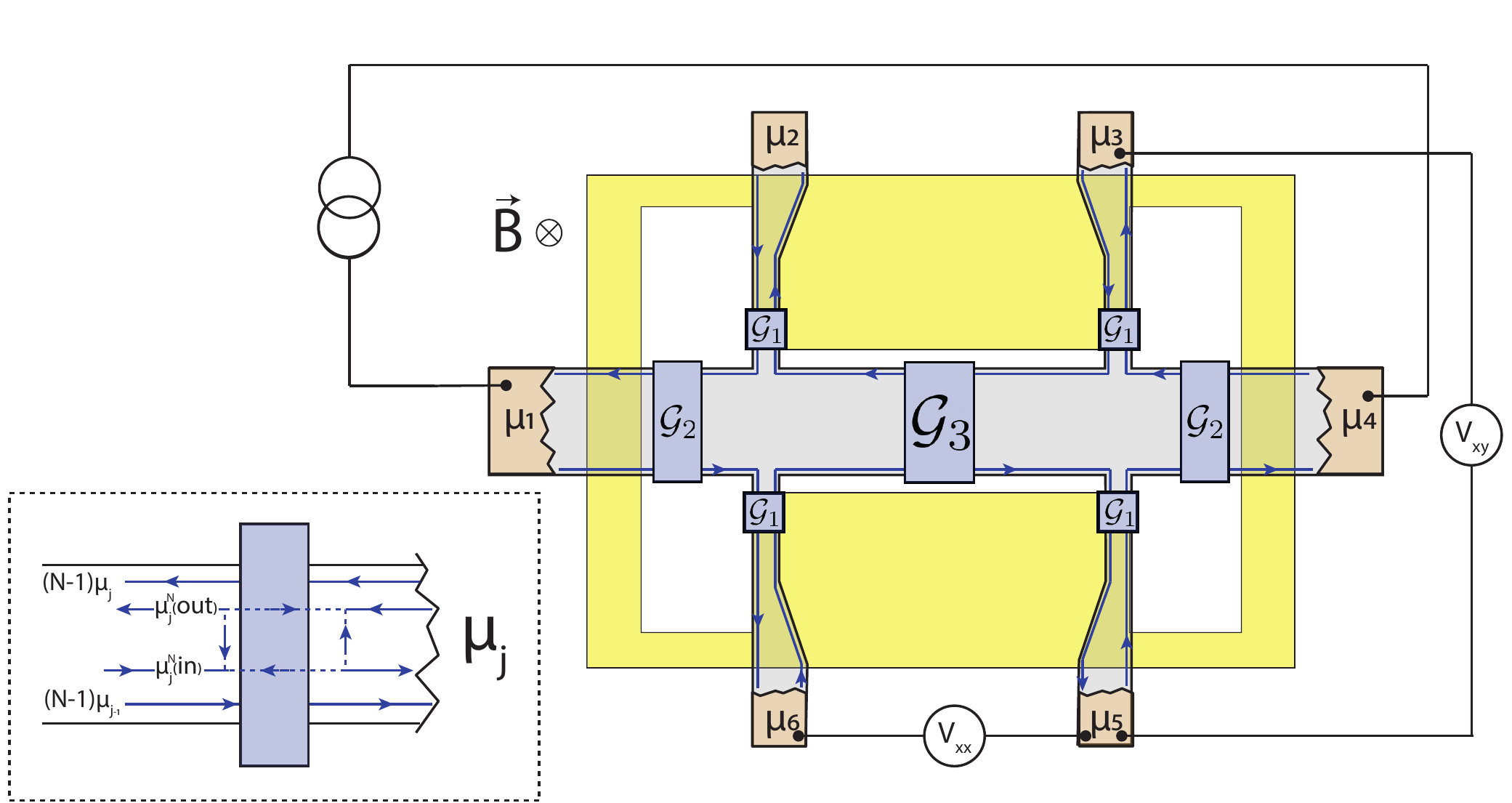}
  \caption{Schematic representation of the setting used to derive the values of the resistances}
  \label{fig:scheme_currents_real_probes}
\end{figure}

As discussed in the main text, the Vacuum fluctuations induce a cavity-assisted long range hopping between states that break the topological protection of the quantum Hall effect. To estimate the effect of these fluctuations onto the quantum Hall transport, we adapted a model of transport initially developed to compute the transport in the magnetic field region in-between quantum Hall plateaux~\cite{mceuen1990new,szafer1991network}. The key point of this model is to assume that backscattering is dominated by the innermost edge channel which is considered to be completely decoupled from the other channels that still flow in a dissipationless manner. 
As shown schematically in Fig.~\ref{fig:scheme_currents_real_probes}, in this model each section of the conductor $j$ is assumed to transmit only a fraction $t_j$ of the inner most edge state $N$, the fraction $1-t_j$ being reflected. Applying the B\"uttiker multiprobe formula~\cite{buttiker1988absence} to the contact yields for the current $I_j$ in each lead $j$ 
\begin{equation}
I_j = \left ( \frac{e}{h} \right) \left [ (N-1) (\mu_j - \mu_{j-1}) - \mu_j^{(\textit{in})} +\mu_j^{(\textit{out})} \right ]
    \label{equ:buttikeroneprobe}
\end{equation}
where the chemical potential $\mu_j^{\text{in}}$ and $\mu_j^{\text{out}}$ represent the chemical potential of a fictitious probe that would be attached to the inner most channel only. Current conservation at the barrier, furthermore, requires
\begin{equation}
    \mu_j^{\text{out}} = \mu_j t_j + \mu_j^{\text{in}} (1-t_j).
    \label{equ:currentconservation}
\end{equation}

The geometry of our sample has been assumed to be well described by the 6 contacts Hall bar described schematically in Fig.~\ref{fig:scheme_currents_real_probes}. As a result, for each pair of contacts an additional equation will link the "outgoing"  chemical potential of the inner edge state of contact $\mu_j^{\text{out}}$ with the incoming potential $\mu_j^{\text{in}}$. 

The scattering induced by the vacuum fluctuations is represented by the transmission coefficients $t_j$ and is assumed to be the same and equal to $t_1$ for the two current probes ($j = 1,4$) and for all the voltage probes $j = 2,3,5,6$. The center section of the conductor is assumed to have a transmission $t_3$. These transmissions are assumed to be related to a vacuum-induced resistivity $\rho_{xx}^{\nu}$ of the innermost edge state $N=\nu$ by
\begin{equation}
    t_j = \frac{1}{1+ \rho_{xx}^{\nu} \mathcal{G}_j} \, ,
\end{equation}
where $\mathcal{G}_j$ is the geometrical aspect ratio $L_j/w_j$ corresponding to the relevant probe {\em describing the interaction with the vacuum field}. 

The above equations can be cast in a $18 \times 18$ matrix $\bar{M}$ relating the vector of currents $\bar{I}$ to the vector of chemical potentials $\bar{\mu}$. Assuming all currents are zero except for the current incoming at contact $1$ and exiting at contact $4$, the system
\begin{equation}
    \bar{I} = \bar{M}\cdot \bar{\mu}
    \label{equ:Hallbarsystem}
\end{equation}
can be solved. The geometrical factors are reported in Table \ref{tab:geometrical-factors}. 
\begin{table}[H]
    \centering
    \begin{tabular}{c c c c}
    \toprule
         Resonator &$\mathcal{G}_1$ &$\mathcal{G}_2$ &$\mathcal{G}_3$ \\
         \midrule
         140GHz (Fig. 1) & 1 & 0.2 &4\\
         cavity 1 180GHz (Fig.6)& 1 & 0.2 & 3.25\\
         cavity 2 180GHz (Fig.6)& 1 & 0.65 & 3.25\\
         cavity 3 180GHz (Fig.6)& 1 & 2 & 3.25\\
         \bottomrule

    \end{tabular}
    \caption{Geometrical parameters used for the different resonators}
    \label{tab:geometrical-factors}
\end{table}

The  system of equations (\ref{equ:Hallbarsystem}) can be solved by setting the chemical potential of one probe to zero. The value of  $\rho_{xx}^{\nu}$ is then the result of matching the computed and experimental values of $R_{xx}$. The correct solution is found by minimizing the error on the computed value of $R_{xy}$, which is then treated as a measurement of the quality of the fit. 
\subsection*{Sample D170608A-Temperature study}

As discussed in the main text, the measurements of the transverse resistivity were repeated varying the temperature.
By applying heat to the mK plate of the dilution refrigerator and controlling the measured temperature through a PID loop back feeding the readings of a resistor bridge temperature sensor, we were able to stabilize the temperature and vary it in the range from $12$ to $700$ mK.

\begin{figure}[H]
 \includegraphics[width=\textwidth]{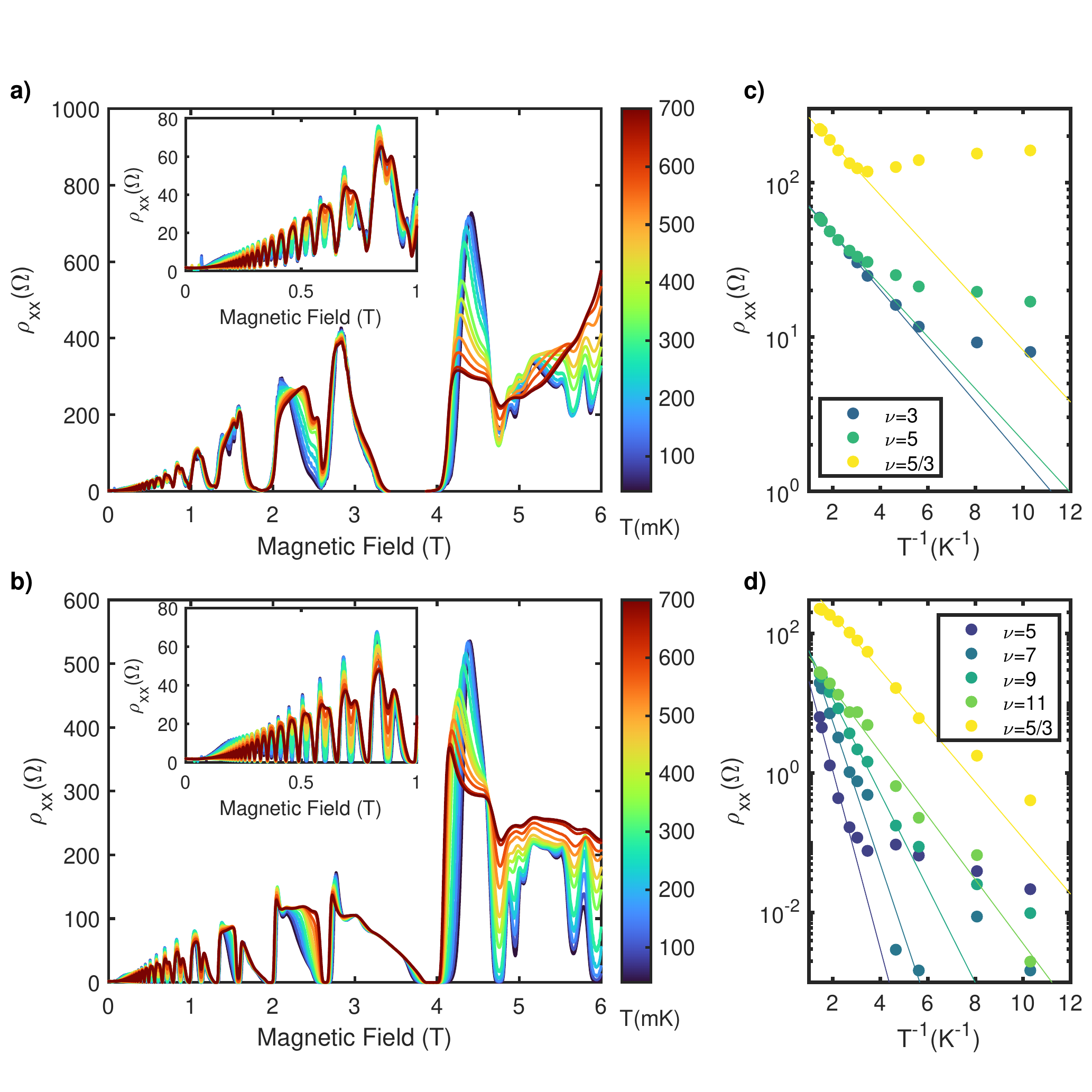}
  \caption{(a): temperature sweep of longitudinal resistance as a function of magnetic field for the complementary split ring resonator with frequency $140$ GHz. (b) temperature sweep of the longitudinal resistance as a function of magnetic field for the reference Hall bar. (c): Arrhenius plot corresponding to (a). (d): Arrhenius plot corresponding to (b).}
  \label{fig:temperature_dependence}
\end{figure}

The temperature study allows us to determine the electronic temperature, which for our system is estimated to be approximately 50mK.
Furthermore we can exploit the temperature dependence of the transport features to determine the activation energies.
The procedure is the following: the values of the minima of $\rho_{xx}$ for different temperatures are extracted.
These values are plotted in log scale as a function of the inverse temperature. From the resulting plot, also known as Arrhenius plot, a linear fit is performed from the region where the thermally activated resistivities vary linearly in log scale. The slope of this fit provides the activation energy of the considered transport feature.
In Fig.\ref{fig:temperature_dependence}, we show the data set used to derive the values of the activation energies shown in Fig. 3 of the main text.

\subsection*{Complementary split-ring resonator design}

\begin{figure}[H]
  \includegraphics[width=\textwidth]{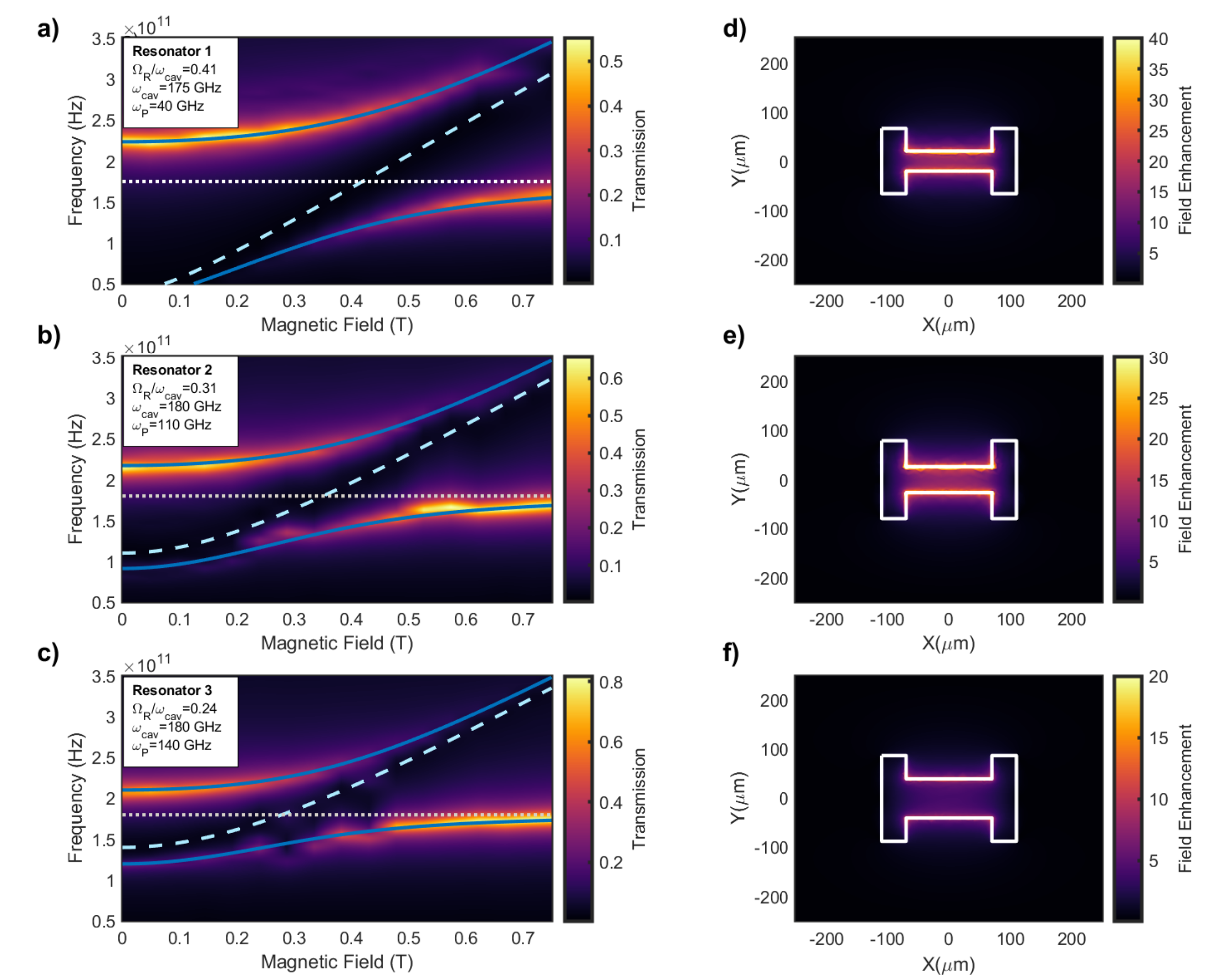}
  \caption{Left: Color plot of the simulated transmission as a function of magnetic field and frequency exhibiting polaritonic dispersions due to strong light-matter coupling for the three cavities. Right: field enhancement for the three cavities. The boundaries of the complementary resonators are highlighted by white lines. }
  \label{fig:simulations_hwga}
\end{figure}

As discussed in the main text, we designed three cavities with fundamental mode frequency around $180$ GHz so that they would be as similar as possible in terms of outer dimensions (the length of the three cavities), but with different normalized coupling strength.
The normalized coupling strength depends on the effective cavity volume and on the number of coupled electrons as $\frac{\Omega_{\text R}}{\omega_{\text{cav}}}\propto\sqrt{\frac{N_{\text{2DEG}}}{V_{\text{cav}}}}$\cite{hagenmuller2010ultrastrong}. Considering that the Hall bar has the same length and width for all the resonators, and hence the same number of electrons $N_{\text{2DEG}}$, scaling the capacitive gap will modulate the coupling strength of the resonator, by changing the fraction of electrons filling the area where the electric field is localized and, in second order by changing the effective transverse length of the electric field. All the simulations of Fig. \ref{fig:simulations_hwga} are performed using CST Microwave Studio.
The resonator is modeled using the standard lossy metal gold from the material library.
The substrate is a block of GaAs. The 2DEG stripe is modeled using a gyrotropic material, with bias in the direction perpendicular to the surface. An effective layer thickness was used in order to reduce computational cost.

This method tends to overestimate the coupling strength, but offers a reliable verification of the expected trend for the resonators, which are then measured optically using our THz spectroscopy set-up described below.

\subsection*{THz-TDS Transmission measurements}

\begin{figure}[H]
  \includegraphics[width=\textwidth]{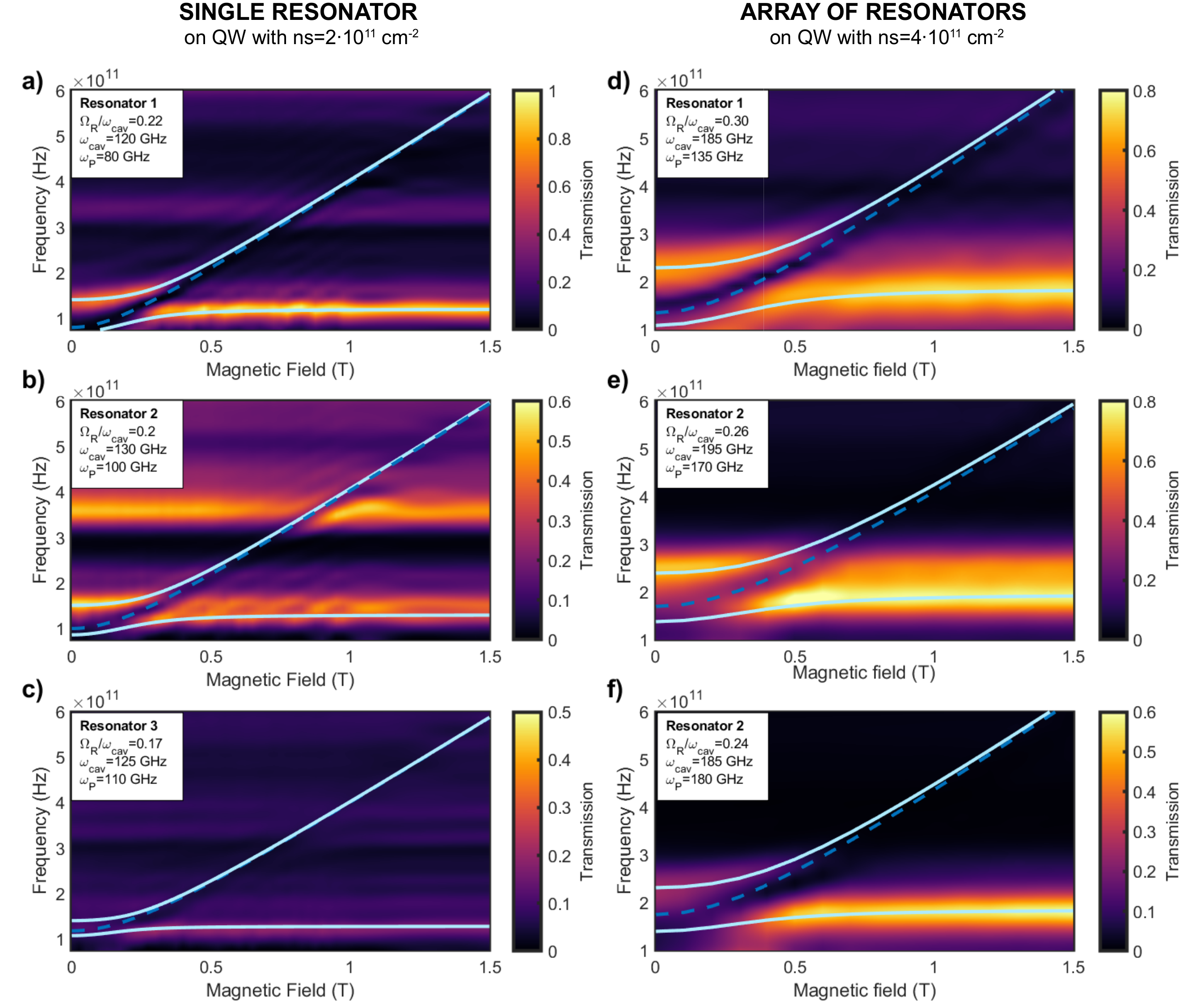}
  \caption{Color plot of the experimentally measured transmission as a function of magnetic field and frequency for the four different samples.}
  \label{fig:sample2_Transmission}
\end{figure}

In Fig.\ref{fig:sample2_Transmission} we present the results of the optical characterization of the resonators described in the previous section.

We show two sets of data. The first one (Fig.\ref{fig:sample2_Transmission}, right) is a set of samples fabricated as arrays. The fits are performed minimizing the RMS considering an Hopfield model where the material part coupled to the cavity mode is a magnetoplasmonic excitation\cite{paravicini2017gate}. 
All the measurements are performed at 3K in a wet cryostat using a Terahertz Time Domain Spectrometer (THz-TDS).
The fabrication of this set of sample was performed on a sample with a triangular well with electronic density under illumination of about $4\times10^{11}cm^{-2}$. It is crucial to notice this, since it is a factor 2 higher than the one of the sample measured in transport.
The expected correction is a factor $\sim\frac{1}{\sqrt{2}}$ in the normalized coupling strength.

The second set of data (Fig.\ref{fig:sample2_Transmission}, left) is an optical characterization of single resonators \cite{singleresonator}
This resonators are fabricated on a triangular quantum well with density $2\times10^{11}cm^{-2}$, same as in the transport measurements.
The technique we adopted to measure the single resonator makes use of two Si immersion lenses. As shown in the reference, the presence of the lenses does not affect the coupling strength significantly, but produces a red-shift of the measured frequencies, due to the change of the effective refractive index of the cavities.

Fitting the resulting polaritonic dispersions, we get values of normalized coupling strength of $\tilde\Omega_{res}/\omega_{cav}=$0.22, 0.2 and 0.17 for the three cavities.

\subsection*{Sample F150817A}

\begin{figure}[H]
  \includegraphics[width=\textwidth]{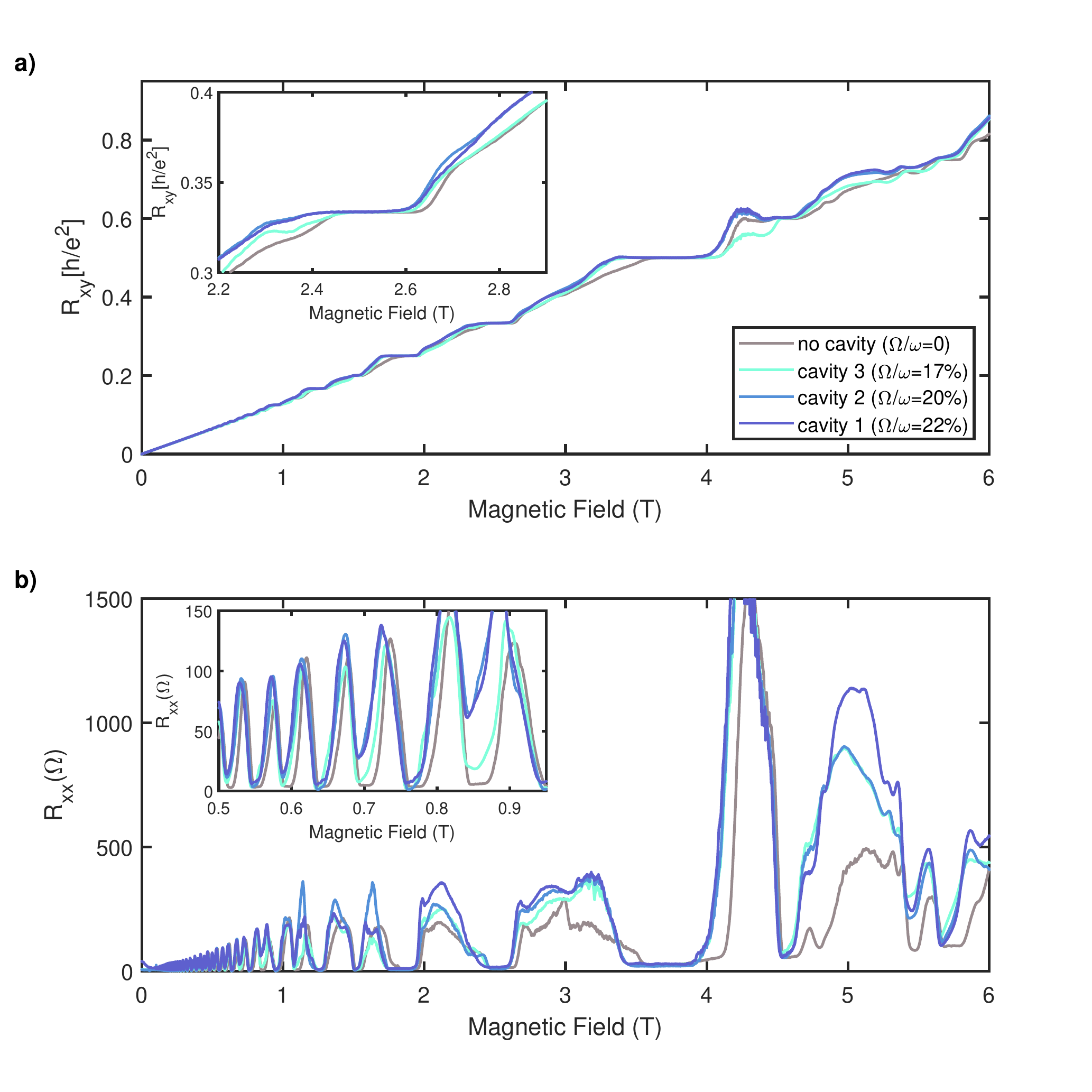}
  \caption{a)transverse resistance measured for the 3 cavities. The inset shows a zoom around filling factor $\nu$=3  b)Longitudinal resistance measured for the 3 cavities. The inset shows a zoom on the minima at low fields. }
  \label{fig:sample2_rxx_and_rxy}
\end{figure}

We report here the full set of data from the sample used for the coupling strength dependence shown in Fig. 6 of the main text.
In Fig. \ref{fig:sample2_rxx_and_rxy} of this Supplementary Material, we show the measured longitudinal and transverse resistances for the three cavities of the scaling study and the reference Hall bar.
In the inset of Fig.\ref{fig:sample2_rxx_and_rxy} a), the zoom at filling factor $\nu=3$ show trend with coupling consistent with the features described in the main text. 

The inset of Fig.\ref{fig:sample2_rxx_and_rxy} b) is a zoom at high filling factors to highlight the deviation from the zero resistance states of the well quantized reference hall bar. The dependence on the coupling strength is evident. These set of data are used for the extraction of the resistivity $\rho^{\nu}_{xx}$ of Fig.6 of the main text.
\begin{figure}[H]
  \includegraphics[width=\textwidth]{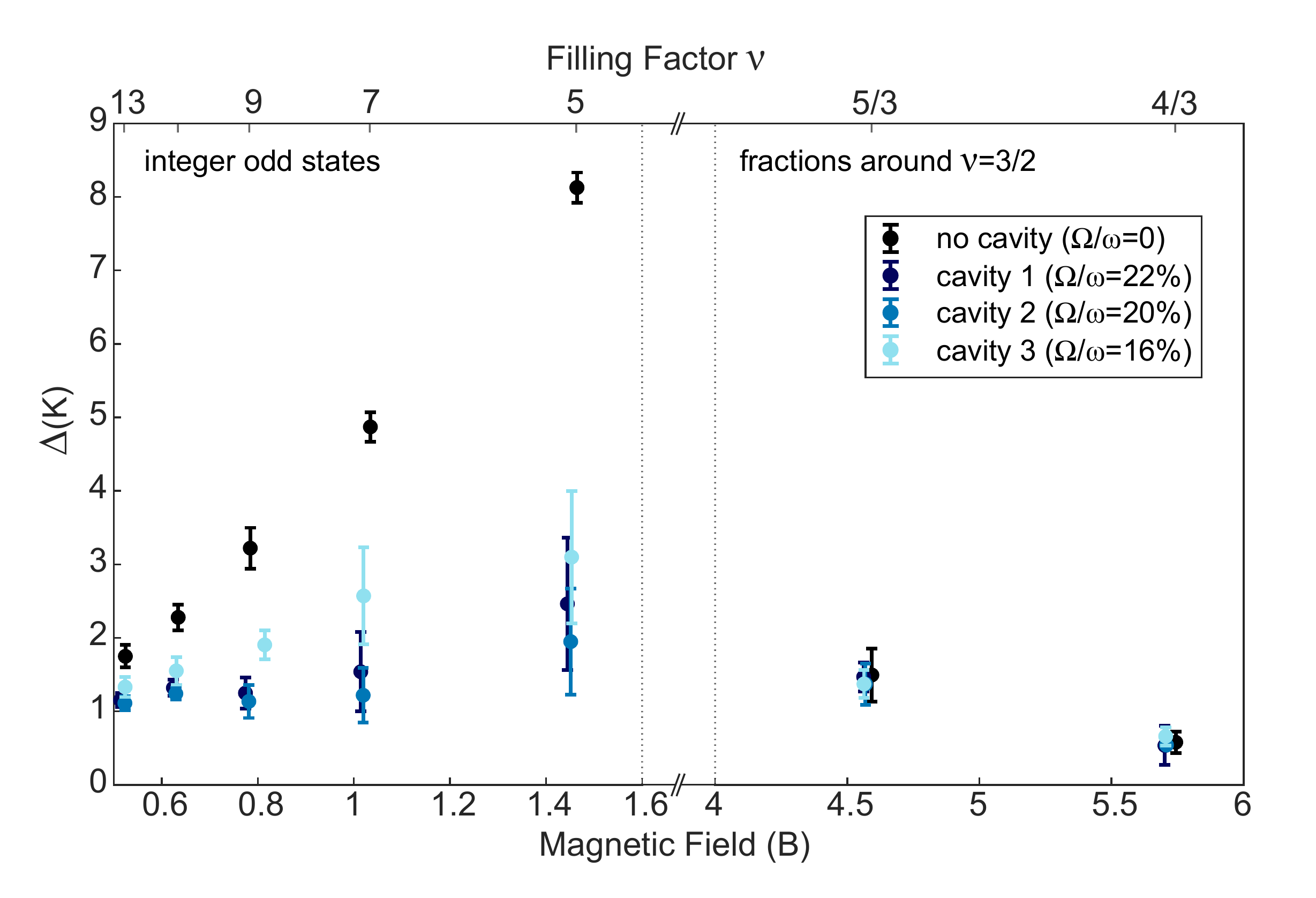}
  \caption{Activation energies (divided by the Boltzman constant) as a function of magnetic field/filling factor for four different samples.
  \label{fig:sample2_activation}}
\end{figure}

\begin{figure}[H]
  \includegraphics[width=\textwidth]{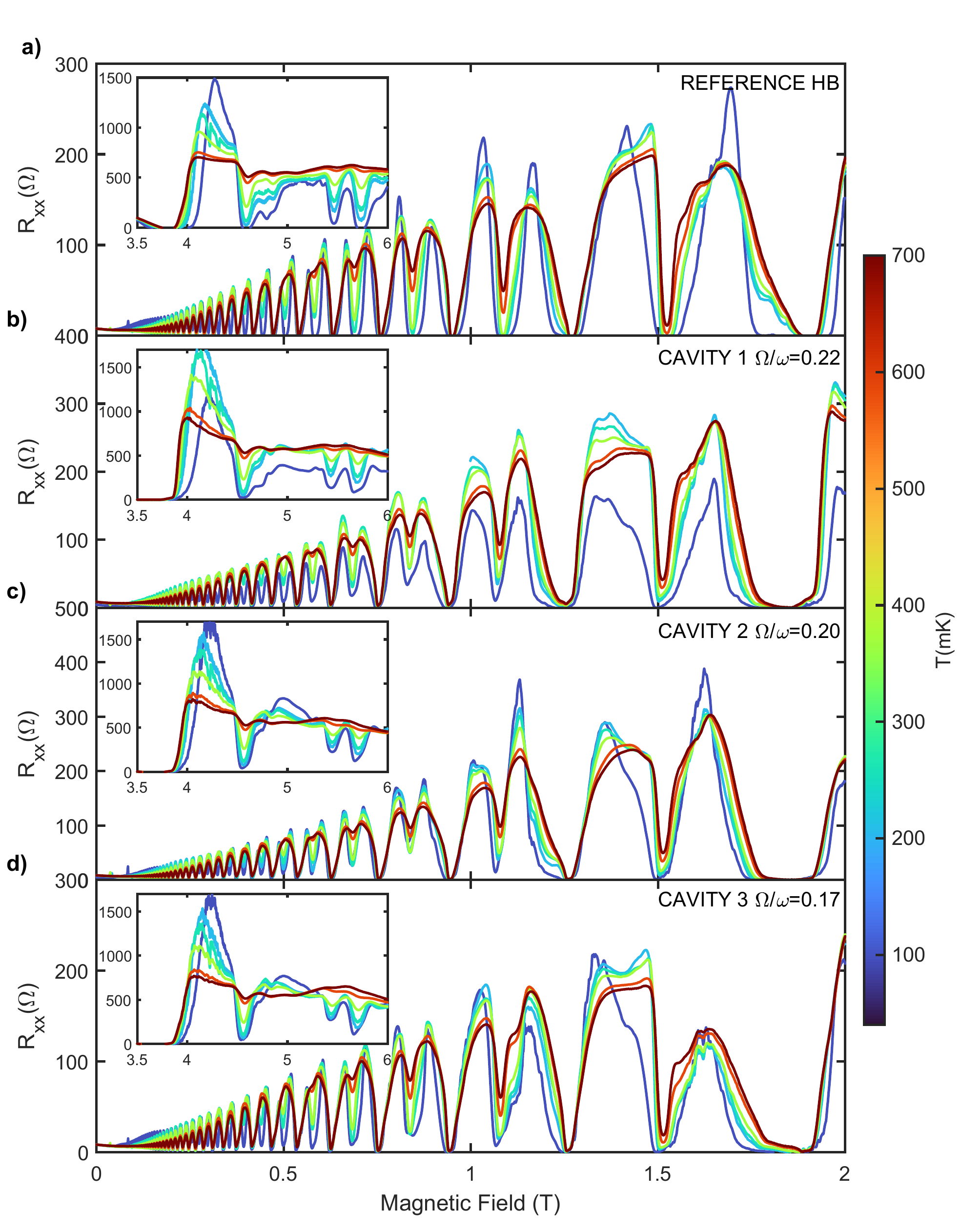}
  \caption{Temperature sweep for different cavities and the reference Hall bar }
  \label{fig:sample2_T_sweep}
\end{figure}

In Fig.\ref{fig:sample2_activation}, we show the thermal dependence study of this sample.
This shows an high degree of consistency with what we discussed for the sample D170608A. In particular, from the inset is noticeable that the activation for the fractional states around $\nu=\frac{3}{2}$ behave very similarly for the cavities and the reference hall bar, while a bigger difference is observed for integer odd states.

The quantitative verification of this statement is provided by the extraction of the activation energies, using the method described in previous sections.
The results are shown in Fig.\ref{fig:sample2_activation}.

\subsection*{Sample D170209B}

\begin{figure}[H]
  \includegraphics[width=\textwidth]{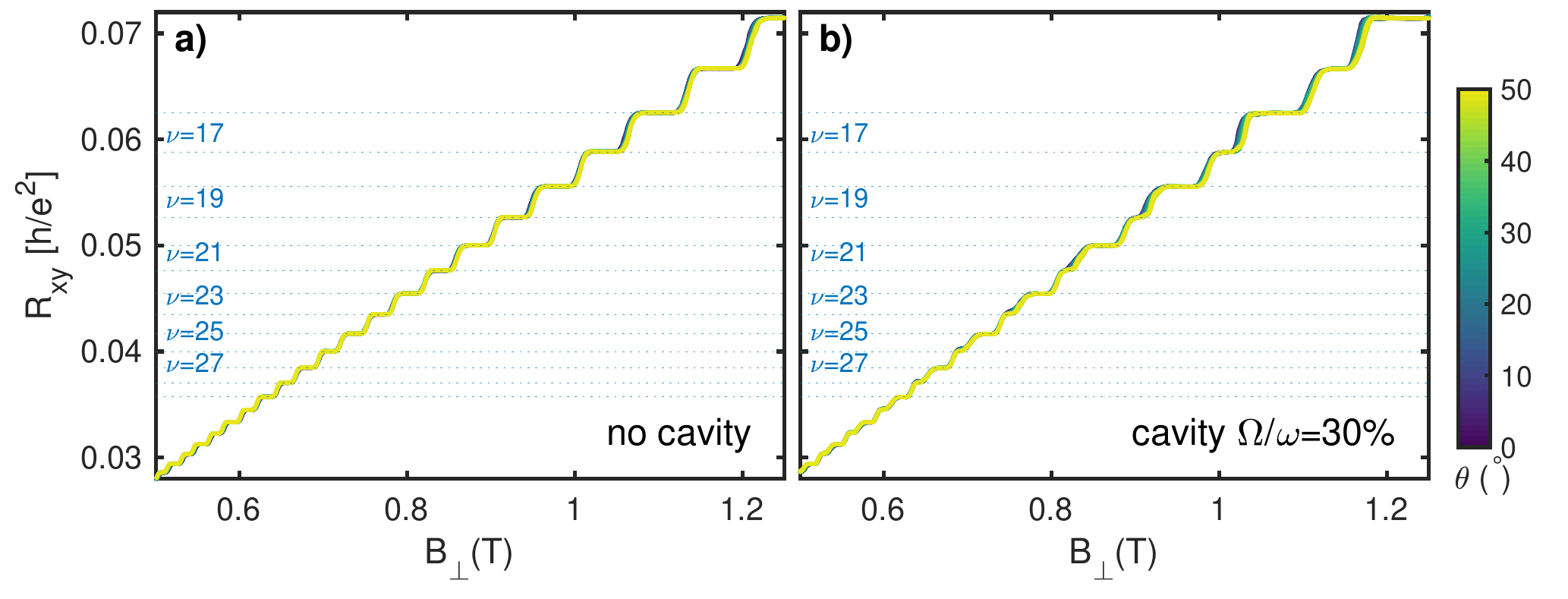}
  \caption{Transverse resistance as a function of the magnetic field component $B_{\perp}$ for sample without(a) and with cavity (b).}
  \label{fig:samplerotation}
\end{figure}

As a complement of the data shown in Fig. 4 of the main text, we present in Fig.\ref{fig:samplerotation} a larger range of magnetic fields, where the effect on the odd plateaus is still visible, even though it is weaker in magnitude.

The limitation to this range of filling factors is mostly given by technical constrains of the magnet used for the study and the density of the measured sample.

\subsection*{Sample EV2124}

\begin{figure}[H]
  \includegraphics[width=\textwidth]{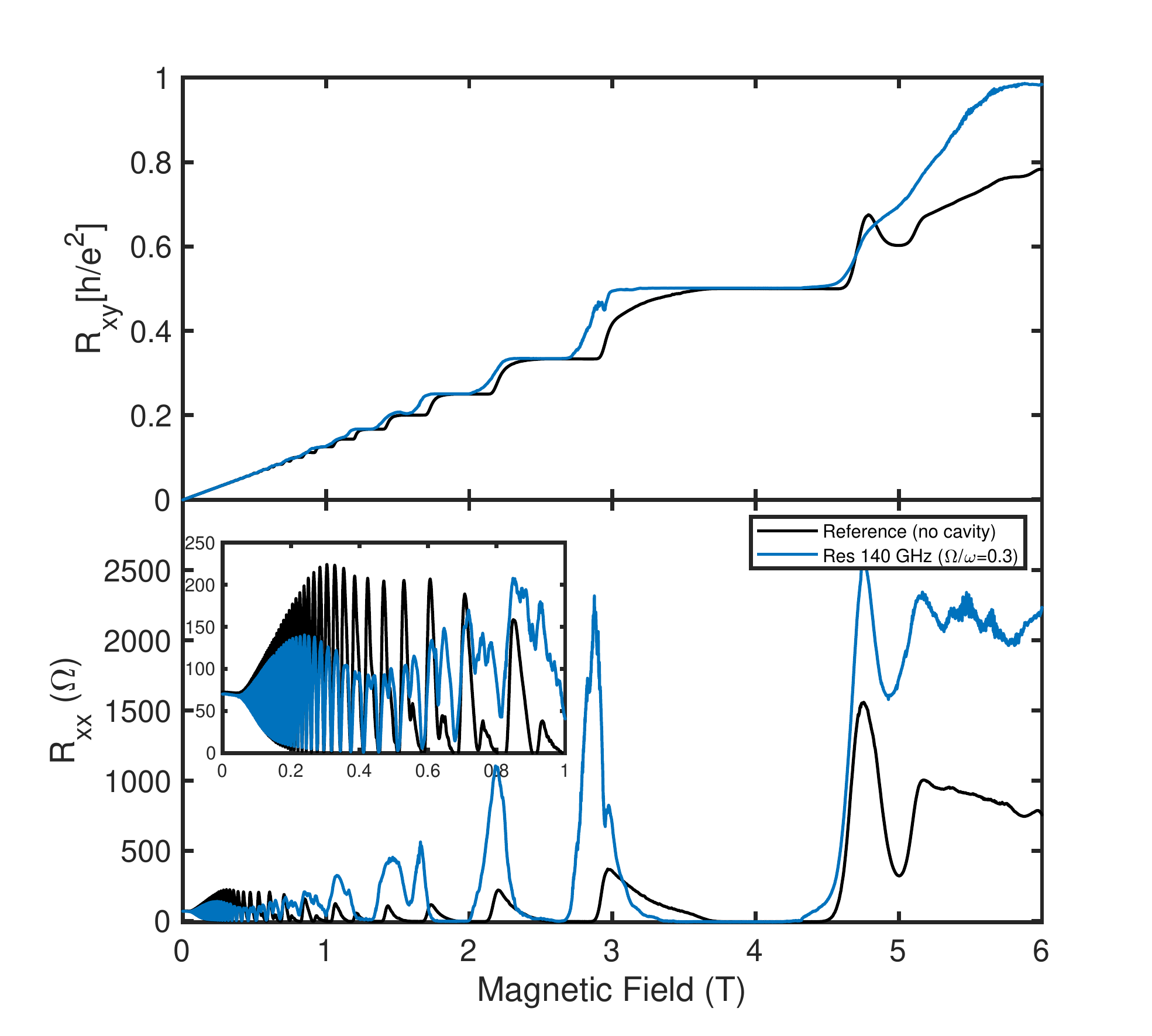}
  \caption{$R_{xx}$ and $R_{xy}$ for sample EV2124.}
  \label{fig:sampleEV}
\end{figure}

The longitudinal $R_{xx}$ and transverse $R_{xy}$ resistances for the sample EV2124 are shown in Fig.\ref{fig:sampleEV}.
The overall behaviour and features are consistent with what is observed in samples fabricated on different quantum wells.

\end{document}